\documentclass[12pt]{article}
\usepackage{amssymb,epsf,graphicx}

\newcommand{\Tr}{\mathop{\rm Tr}\nolimits}

\def\ket#1{|#1 \rangle}
\def\aver#1{\langle\, #1 \,\rangle}

\def\l{\left}
\def\r{\right}

\def \be {\begin{equation}}
\def \ee {\end{equation}}
\def \bea {\begin{eqnarray}}
\def \eea {\end{eqnarray}}
\def \bdm {\begin{displaymath}}
\def \edm {\end{displaymath}}

\def \oo {{\cal O}}

\def \vv {{\cal V}}

\def \sft {string field theory }

\begin{document}
\begin{flushright}
IFT-P.027/2003\\
CTP-MIT-3393\\
{\tt hep-th/0307019}
\end{flushright}

\vskip 2cm

\begin{center}
{\large \bf Yang-Mills Action from Open Superstring Field Theory\\}
\vskip 1cm

{\large Nathan Berkovits}\footnote{E-mail: nberkovi@ift.unesp.br} \\
\vskip 0.5cm
{\it Instituto de F\'{\i}sica Te\'orica, Universidade Estadual Paulista,\\
Rua Pamplona 145, S\~ao Paulo, SP 01405-900, Brasil}
\vskip 1 cm

{\large Martin Schnabl}\footnote{E-mail: schnabl@lns.mit.edu} \\
\vskip 0.5cm
{\it Center for Theoretical Physics, Massachusetts Institute of Technology,\\
Cambridge, MA 02139, USA}
\end{center}

\begin{abstract}
We calculate the effective action for nonabelian gauge bosons up
to quartic order using WZW-like open superstring field theory. After
including level zero and level one contributions, we obtain with 75\%
accuracy the Yang-Mills quartic term. We then prove that the complete
effective action reproduces the exact Yang-Mills quartic term by
analytically performing a summation over the
intermediate massive states.
\end{abstract}

\newpage 
\baselineskip=18pt

\section{Introduction}

Although the conventional way to compute effective actions in string theory
uses two-dimensional sigma model methods, one can also compute
the effective action using string field theory methods by integrating
out the massive fields. This was shown for the open bosonic string
in a recent paper by 
Coletti, Sigalov and Taylor \cite{Coletti:2003ai}, based on earlier
work by Taylor \cite{Taylor:2000ek}. Since 
gauge transformations in string field theory mix the massless and
matter fields, it is interesting to see how the usual low-energy
gauge invariances are recovered after integrating out the massive fields.

In this paper, we generalize the methods of 
\cite{Taylor:2000ek,Coletti:2003ai}
for the superstring using the WZW-like version of open superstring
field theory \cite{Berkovits:1995ab,Berkovits:2001}. 
After including level zero
and level one contributions to the effective action, we find that
the correct Yang-Mills quartic coupling is reproduced with 75\% accuracy. 
We then show 
by performing an explicit summation over all intermediate massive states
that the exact
Yang-Mills quartic coupling is reproduced in both bosonic
string field theory and in WZW-like superstring field theory.
Since there are several alternative proposals \cite{Witten,Yost,Arefeva} 
for open superstring field
theory, it would be interesting to know if any of these alternative
proposals are able to reproduce these results.

In section 2 of this paper, we compute the purely massless contribution
to the effective action, i.e. the contribution before integrating out
massive fields. Unlike the bosonic case, there is already
a quartic coupling in the WZW-like string field theory action
before integrating
out massive fields. However, it reproduces the correct Yang-Mills
quartic coupling with only 50\% accuracy.

In section 3 of this paper, we include the contribution from integrating
out level one massive fields. At this level, the correct Yang-Mills
quartic coupling is reproduced with 75\% accuracy. 

Finally, in section 4 of this paper we prove by an explicit summation
that the correct Yang-Mills quartic coupling is reproduced after
integrating out all massive states. This proof is given both for
Witten's bosonic string field theory and for WZW-like superstring
field theory. The proof confirms an earlier result \cite{Berkovits:1999bs}
that WZW-like
superstring field theory reproduces the correct on-shell four-point
amplitudes without contact-term divergences.

\section{Contribution of massless fields}

In this section, we shall compute the purely massless contribution 
to the effective action.
In WZW-like open superstring field theory, 
the action on a BPS $D$-brane is 
\cite{Berkovits:1995ab,Berkovits:2001}
\be 
S = \frac{1}{2g^2} \l\langle
\l(e^{-\Phi} Q e^{\Phi} \r) \l(e^{-\Phi} \eta_0 e^{\Phi} \r)
- \int_0^1 dt \l(e^{-t\Phi} \partial_t e^{t\Phi} \r) \l\lbrace 
\l(e^{-t\Phi} Q e^{t\Phi} \r), \l(e^{-t\Phi} \eta_0 e^{t\Phi}\r)
\r\rbrace 
\r\rangle
\ee
where $\Phi$ is a Neveu-Schwarz string field in the GSO(+) sector
at picture zero
and ghost-number zero in the large Hilbert space, $Q$ is
the BRST charge
\be
Q = \oint \frac{dz}{2\pi i} \l[c(T_m+T_{\xi\eta} + T_\phi) +
c\partial c b + i \sqrt{\frac{2}{\alpha'}} \eta e^\phi  \psi_\mu
\partial X^\mu - \eta \partial \eta e^{2\phi} b\r],
\ee 
and we use the normalization
$\aver{\xi(z) c\partial c\partial^2 c(w) e^{-2\phi(y)}}=2$.
\footnote{ 
Since we will only compute the
bosonic contribution to the effective action, we can use
the $d=10$ Neveu-Schwarz version of the WZW-like action. However, one could 
easily
compute the fermionic contribution to the effective action using the
$d=4$ supersymmetric version of the action \cite{Berkovitsramond} as
in \cite{Barosiramond}.}
For practical calculations, it is convenient to use the 
somewhat simplified form of the action \cite{Berkovits:2000hf}
\be
S = \frac{1}{g^2} \sum_{M,N =0}^\infty \frac{1}
{(M+N+2)!} \l(\begin{array}{c} M+N \\ M \end{array} \r)
(-1)^N  \l\langle
\l(Q \Phi\r) \Phi^M \l(\eta_0 \Phi \r) \Phi^N \r\rangle .
\ee

At the massless level, there are two string fields 
\bea
\Phi_A &=& \int \!d^dk \, A_\mu(k) \vv_A^\mu(k), \nonumber\\
\Phi_B &=& \int \!d^dk \, B(k) \vv_B(k),
\eea
where
\bea
\vv_A^\mu(k) &=& c \,\xi \, e^{-\phi} \psi^\mu e^{i k X}, \nonumber\\
\vv_B(k) &=& c\partial c \,\xi \partial \xi \,e^{-2\phi} e^{ik X}
\eea
are the massless vertex operators for the gluon and the auxiliary 
Nakanishi-Lautrup field.
One can easily compute that 
\bea
Q\vv_A^\mu(k) &=& 
-\alpha' k^2 c \partial c \, \xi \, e^{-\phi} \psi^\mu e^{i k X} \nonumber\\
&& +\sqrt{2\alpha'} c \l( k \cdot \psi\, \psi^\mu + {\frac{i}{\alpha'}}
\partial X + (:\eta
\xi: + \partial\phi) k^\mu \r) e^{i k X} \nonumber\\
&& + \eta e^\phi \psi^\mu e^{i k X}
\eea
and
\bea
Q\vv_B(k) &=& \sqrt{2\alpha'} c\partial c \,\xi \, k.\psi
\,e^{-\phi} e^{ik X}
\nonumber\\
&& + \l(-\partial c + 2 c (:\eta \xi: + \partial\phi)\r) e^{i k X}.
\eea

The contribution of these fields to the 
\sft action is easily computed to be
\bea
S &=& \frac{1}{g^2}\int \!d^dx \Tr \l[ 
\frac{\alpha'}{2} A_\mu
\partial^2
A^\mu 
 - i\sqrt{2\alpha'} B \, \partial^\mu A_\mu 
+ B^2 \r. \nonumber\\
&& \l.
-\frac{i\sqrt{2\alpha'}}{2} 
\partial_\mu A_\nu \l[A^\mu , A^\nu \r]
 +\frac{1}{8}
A_\mu A_\nu A^\mu A^\nu - \frac{1}{2} A_\mu A_\nu A^\nu A^\mu \r] .
\eea
Note that the cubic coupling of $A_\mu A^\mu B$ 
vanishes in the superstring just as it does in the bosonic string.
In addition,
the coupling $A_\mu\, \partial^\mu\! B B$ also vanishes for the superstring.

After integrating out the auxiliary field $B$ 
we find 
\be\label{truncatedaction}
S = 
S_{YM} - \frac{1}{4g^2} \int \! d^dx \Tr \l[ \frac{1}{2} A_\mu A_\nu A^\mu
A^\nu +  A_\mu A_\nu A^\nu A^\mu \r],
\ee
where
\be
S_{YM} = - \frac{\alpha'}{4g^2} \int \! d^dx \Tr \l( \partial_\mu A_\nu -
\partial_\nu A_\mu + \frac{i}{\sqrt{2\alpha'}} \l[A_\mu, A_\nu\r]\r)^2.
\ee
So the quartic terms in the effective
action at level zero,
\be\label{quarticzero}
S_{\mathrm{zero} } = 
\frac{1}{g^2} \int \! d^dx \Tr \l[ \frac{1}{8} A_\mu A_\nu A^\mu
A^\nu - \frac{1}{2} A_\mu A_\nu A^\nu A^\mu \r],
\ee
are off by approximately 50\% from the gauge-invariant quartic terms
which are
\be\label{quarticcorrect}
S_{\mathrm{exact} } = 
\frac{1}{g^2} \int \! d^dx \Tr \l[ \frac{1}{4} A_\mu A_\nu A^\mu
A^\nu - \frac{1}{4} A_\mu A_\nu A^\nu A^\mu \r].
\ee

\section{Contribution of level one massive fields}

To compute the contribution of massive fields to the quartic Yang-Mills
term,
one only needs to consider massive vertex operators 
$\vv_C(k)$ for which the cubic vertex 
\bea
\langle \vv_C,  Q \vv_A * \eta_0 \vv_A \rangle
\eea
is non-vanishing at zero momentum.
Furthermore, we shall only consider massive vertex operators in
Siegel gauge, i.e. those vertex operators satisfying $b_0 \vv_C=0$.

One can easily show that the only two massive vertex operators at level one
which satisfy the above requirements are
\be
\vv_C^{\mu\nu}(k) = c \,\xi \, e^{-\phi} \psi^\mu \partial X^\nu e^{i k X}
\ee
and 
\be
\vv_D(k) = c \,\partial^2 c\,\xi \,\partial\xi  e^{-2\phi} e^{i k X}.
\ee
However, the vertex operator $\vv_D$ has vanishing quadratic
coupling with itself and with $\vv_C^{\mu\nu}$ and therefore does
not contribute to the quartic Yang-Mills term. So the only contribution
in Siegel gauge to the quartic term from level one massive states comes
from the state $C_{\mu\nu}(k)$ which couples to $\vv_C^{\mu\nu}$.

Computing the quadratic and cubic interactions for $\vv_C^{\mu\nu}$, one
finds  
\be
 S = \frac{1}{g^2} \int \! d^dx 
\Tr \l[ \frac{\alpha'}{4} C_{\mu\nu}  C^{\mu\nu} 
- i \frac{\sqrt{2\alpha'}}{3\sqrt{3}} C^{\mu\nu} 
\lbrace A_\mu, A_\nu \rbrace \r].
\ee
So after integrating out the $C_{\mu\nu}$ field, one finds that the
level one massive contribution in Siegel gauge to the quartic term is
\be
S_{\mathrm{one} } = 
\frac{1}{g^2} \int \! d^dx \Tr \l[\frac{4}{27}
 A_\mu A_\nu A^\mu A^\nu +
\frac{4}{27} A_\mu A_\nu A^\nu A^\mu \r]
\ee
Summing this level one contribution to the level zero contribution of
(\ref{quarticzero}), one finds
\be
S_{\mathrm{zero} } +
S_{\mathrm{one} } 
 = 
\frac{1}{g^2} \int \! d^dx \Tr \l[\frac{59}{216}
 A_\mu A_\nu A^\mu A^\nu -
\frac{19}{54} A_\mu A_\nu A^\nu A^\mu \r],
\ee
which is off from the correct quartic coupling of (\ref{quarticcorrect})
by approximately 25\%. 

For comparison,
we show in Table 1
the numerical values of the coefficients 
of the quartic term at zeroth and first levels together 
with their exact values.
The results indicate good convergence properties similar to those
of the bosonic string \cite{Coletti:2003ai}.
\begin{table}\label{table-quartic}
\begin{center}
\begin{tabular}{|l|r|r|}
\hline
& {\footnotesize $A_\mu A_\nu A^\mu A^\nu$} 
&  {\footnotesize $A_\mu A^\mu A_\nu A^\nu$}
\\ \hline
level 0 & 0.50 & -2.00
\\ \hline
level 1 & 1.09 & -1.41
\\ \hline
exact & 1.00 & -1.00
\\ \hline
\end{tabular}
\end{center}
\caption{\small Coefficients of the quartic term 
for the gauge field normalized by a factor $4g^2$.} 
\end{table}

\section{Analytic calculation of the exact quartic term}

In this section we compute the exact Yang-Mills
quartic term in the effective action
by analytically performing the sum over all intermediate
massive states. This is done first for Witten's bosonic string
field theory and then for the WZW-like
superstring field theory. Note that we are able to avoid
numerical computations like that
of \cite{Coletti:2003ai,Taylor:2000ek} because we are only interested
in the zero momentum contribution of the gauge fields.

\subsection{Quartic term for the bosonic string}

To find the massive contribution to the Yang-Mills quartic
term at zero momentum, consider the
string field 
\be
\Phi = R+ \Phi_A 
\ee
where $R$ is the massive part of the string field and
\be
\Phi_A = 
i \sqrt{\frac{2}{\alpha'}}\, A_\mu c\partial X^\mu 
\ee
is the zero-momentum part of the gauge boson.
The contribution of $R$ to Witten's action is
\bea
S &=& 
 -\frac{1}{g^2}\l[\frac{1}{2} \aver{R, Q R} + \aver{R,\Phi_A*\Phi_A} + 
 \aver{\Phi_A,R*R} +\frac{1}{3} \aver{R, R*R}\r].
\eea
Shifting
\be\label{R-sol}
R \to R -\frac{b_0}{L_0} \Phi_A*\Phi_A 
\ee
to eliminate the terms linear in $R$,
one learns that
the massive contribution to the quartic Yang-Mills term is
\be\label{quartic-general}
S^{(4)} = \frac{1}{2g^2} \aver{\frac{b_0}{L_0} \Phi_A*\Phi_A, Q \frac{b_0}{L_0}
\Phi_A*\Phi_A}.
\ee

This term can be conveniently expressed in the operator formalism
as\footnote{By inserting a complete set of states on the left and right hand
sides of $b_0/L_0$, we can make contact with the calculation of
\cite{Coletti:2003ai}, albeit in a different language.}
\bea\label{S4bos-1}
S^{(4)} &=&  \frac{1}{2g^2} \l(\frac{2}{\alpha'}\r)^2 \int \!d^dx \, \Tr(A_\mu
A_\nu A_\rho A_\sigma) \nonumber\\
&&
 \aver{c\partial X^\mu(-\sqrt{3}) \, c\partial X^\nu(\sqrt{3}) \, U_3 \frac{b_0}{L_0}
U_3^\dagger \, c\partial X^\rho\l(\frac{1}{\sqrt{3}}\r) \, c\partial
X^\sigma\l(-\frac{1}{\sqrt{3}}\r)}
\eea
where $U_r$ is a representation of the conformal map $f_r(z)= \tan\l(\frac{2}{r}\arctan z\r)$
on the Hilbert space 
of the string, i.e. $U_r \oo(z) U_r^{-1} =
 f_r \circ \oo(z) = (f_r')^d \oo(f_r(z))$.
(The last equality is true only for primary operators of dimension $d$.) 
As discussed in
\cite{Rastelli:2000iu,Schnabl:2002gg}, 
the $U_r$
operators can be written down explicitly 
as exponentials in the Virasoro operators.

Since the state $c\partial X^\rho\l(\frac{1}{\sqrt{3}}\r) \,
c\partial X^\sigma\l(-\frac{1}{\sqrt{3}}\r)\ket{0}$ 
is a BRST exact state, we
can freely commute the propagator $b_0/L_0$ through the $U$ operators. Then
using the identity \cite{Schnabl:2002gg}
\be
U_3 U_3^\dagger  = U_{\frac{8}{3}}^\dagger  U_{\frac{8}{3}}
\ee
we can commute $U$ to the right and $U^\dagger$ to the left
with the help of
\be
U_{\frac{8}{3}} c\partial X^\sigma\l(\pm \frac{1}{\sqrt{3}}\r)
U_{\frac{8}{3}}^{-1} = c\partial X^\sigma(\pm a),
\ee
where
\be
a = \tan \frac{\pi}{8} = \sqrt{2}-1.
\ee
Now we can easily evaluate the correlator in (\ref{S4bos-1}) as
\bea\label{correlator-bos}
 && \aver{c\partial X^\mu(-\frac{1}{a}) \, 
c\partial X^\nu(\frac{1}{a}) \, \frac{b_0}{L_0} \,
 c\partial X^\rho(a) \, c\partial X^\sigma(-a)} =
 \nonumber\\
 && \quad =
 -\l(\frac{\alpha'}{2}\r)^2 4 \int_0^\infty dt \, e^{-t} \l(e^{-2t} a^2 -
 \frac{1}{a^2} \r) \l( \frac{e^{2t}}{16} \eta^{\mu\nu} \eta^{\rho\sigma} +
\frac{ \eta^{\mu\rho} \eta^{\nu\sigma} }{\l(\frac{1}{a}+e^{-t} a\r)^4} + \frac{
\eta^{\mu\sigma} \eta^{\nu\rho} }{\l(\frac{1}{a}-e^{-t} a\r)^4} \r) \nonumber\\
 && \quad =
 \l(\frac{\alpha'}{2}\r)^2 \l(-\frac{3}{2} \eta^{\mu\nu} \eta^{\rho\sigma}
+\frac{1}{2}\eta^{\mu\rho} \eta^{\nu\sigma} + \eta^{\mu\sigma} \eta^{\nu\rho}
\r)
\eea
where we used
\be
\frac{b_0}{L_0} = \int_0^\infty dt \, b_0 \, e^{- t L_0}.
\ee
Note that
all the terms in the integral are convergent but one, which we defined by
analytic continuation to be
\be\label{tach-trick}
\int_0^\infty dt \, e^{t} = -1.
\ee
This term arises due to the open string tachyon propagating as an intermediate
state and, for this state, the exponential representation of the propagator is
invalid. So one could either treat the tachyon separately 
or use (\ref{tach-trick})
as a convenient shortcut.

Plugging the result for the correlator (\ref{correlator-bos}) into
(\ref{S4bos-1}) we find 
\be
S^{(4)} =  -\frac{1}{4g^2} \int \! d^dx \Tr \l( A_\mu A^\mu A_\nu A^\nu
- A_\mu A_\nu A^\mu A^\nu \r)
\ee
which is the correct quartic term for the Yang-Mills action
\be
S_{eff} = - \frac{\alpha'}{4g^2} \int \! d^dx \Tr \l( \partial_\mu A_\nu -
\partial_\nu A_\mu - \frac{i}{\sqrt{2\alpha'}} \l[A_\mu, A_\nu\r]\r)^2.
\ee

\subsection{Quartic term for the superstring}

In this section we generalize the analytic calculation of the quartic term to
the superstring using the WZW-like action. Substituting the string
field
\be
\Phi = R+ \Phi_A 
\ee
into the WZW-like action where 
$R$ is the massive part of the string field and
\be
\Phi_A = 
A_\mu \, c \,\xi e^{-\phi}\psi^\mu
\ee
is the zero-momentum part of the gauge boson, one obtains
\bea
S &=& 
- \frac{1}{2 g^2}\aver{\eta_0 R, Q R} -\frac{1}{2g^2}
 \aver{\eta_0 R,(Q\Phi_A)*\Phi_A
-\Phi_A *(Q\Phi_A)} + \cdots
\eea
where $\ldots$ are higher-order terms in $R$ which will not contribute
to the quartic Yang-Mills term.
Shifting
\be\label{R-sol-susy}
R \to R+ 
\frac{1}{2} \frac{b_0}{L_0} \l(\Phi_A* Q\Phi_A - Q \Phi_A*\Phi_A\r)  
\ee
to eliminate the terms linear in $R$,
one learns that
the massive contribution to the quartic Yang-Mills term for the
superstring is 
\be\label{quartic-general-susy}
S^{(4)} = -\frac{1}{2g^2} \frac{1}{4} \aver{Q \frac{b_0}{L_0} (\Phi_A*
Q\Phi_A - Q \Phi_A*\Phi_A), \eta_0 \frac{b_0}{L_0} (\Phi_A* Q\Phi_A - Q
\Phi_A*\Phi_A)}.
\ee

Simple evaluation as in the
bosonic case gives
\bea\label{S4bos-1-susy}
S^{(4)} &=& \frac{1}{4g^2 \alpha'}  \int \!d^dx \, \Tr(A_\mu
A_\nu A_\rho A_\sigma) \nonumber\\
&&
 \aver{\l(c\xi e^{-\phi} \psi^\mu(-\sqrt{3}) \, c\partial X^\nu(\sqrt{3})
- c\partial X^\mu(-\sqrt{3}) \, c\xi e^{-\phi} \psi^\nu(\sqrt{3}) \r)
  \, U_3 \frac{b_0}{L_0}
U_3^\dagger
\nonumber\\
&& \l(c e^{-\phi} \psi^\rho\l(\frac{1}{\sqrt{3}}\r) \, c\partial
X^\sigma\l(-\frac{1}{\sqrt{3}}\r) + c\partial X^\rho\l(\frac{1}{\sqrt{3}}\r)
\,c e^{-\phi} \psi^\sigma\l(-\frac{1}{\sqrt{3}}\r) \r)}
\nonumber\\
&=& -\frac{1}{g^2}  \int \!d^dx \, \Tr(A_\mu A_\nu A_\rho A_\sigma)
\nonumber\\
&& \int_0^\infty dt \, e^{-t} \l(e^{-2t} a^2 -
 \frac{1}{a^2} \r) \l(
\frac{ \eta^{\mu\rho} \eta^{\nu\sigma} }{\l(\frac{1}{a}+e^{-t} a\r)^4} + \frac{
\eta^{\mu\sigma} \eta^{\nu\rho} }{\l(\frac{1}{a}-e^{-t} a\r)^4} \r)
\nonumber\\
&=& \frac{1}{4g^2} \int \! d^dx \Tr \l[ \frac{1}{2} A_\mu A_\nu A^\mu A^\nu +
A_\mu A_\nu A^\nu A^\mu \r],
\eea
which cancels precisely the unwanted term in (\ref{truncatedaction}). This
cancellation is guaranteed since we are effectively
calculating the 
on-shell string diagram for the scattering of four gauge bosons at
zero momentum. But the WZW-like \sft was proven in
\cite{Berkovits:1999bs} to reproduce the correct on-shell four-point
amplitudes. So the calculation here can be regarded as an independent
check for this proof.

\section*{Acknowledgments}
We would like to thank E.~Coletti, 
I.~Ellwood, I.~Sigalov and W.~Taylor for useful 
discussions.  The work of M.S. was supported 
in part by DOE contract \#DE-FC02-94ER40818.
N.B. would like to thank MIT for its hospitality where this
work was initiated, and FAPESP grant 99/12763-0, CNPq grant
300256/94-9 and Pronex grant 66.2002/1998-9 for partial
financial support.

\end{document}